\documentclass[sigconf]{acmart}

\usepackage{booktabs} 

\copyrightyear{2018} 
\acmYear{2018} 
\setcopyright{acmlicensed}
\acmConference[PEARC '18]{Practice and Experience in Advanced Research Computing}{July 22--26, 2018}{Pittsburgh, PA, USA}
\acmBooktitle{PEARC '18: Practice and Experience in Advanced Research Computing, July 22--26, 2018, Pittsburgh, PA, USA}
\acmPrice{15.00}
\acmDOI{10.1145/3219104.3219122}
\acmISBN{978-1-4503-6446-1/18/07}

\acmArticle{4}

\begin{document}
\title{Deploying Jupyter Notebooks at scale on XSEDE resources for Science Gateways and workshops}

\author{Andrea Zonca}
\orcid{0000-0001-6841-1058}
\affiliation{%
  \institution{San Diego Supercomputer Center\\University of California, San Diego}
  \streetaddress{9500 Gilman Drive}
  \city{La Jolla} 
  \state{CA} 
  \postcode{92093-0505}
}
\email{zonca@sdsc.edu}

\author{Robert S Sinkovits}
\affiliation{%
  \institution{San Diego Supercomputer Center\\University of California, San Diego}
  \streetaddress{9500 Gilman Drive}
  \city{La Jolla} 
  \state{CA} 
  \postcode{92093-0505}
}
\email{sinkovit@sdsc.edu}

\renewcommand{\shortauthors}{A. Zonca et al.}

\begin{abstract}
Jupyter Notebooks have become a mainstream tool for interactive computing in every field of science. Jupyter Notebooks are suitable as companion applications for Science Gateways, providing more flexibility and post-processing capability to the users. Moreover they are often used in training events and workshops to provide immediate access to a pre-configured interactive computing environment.
The Jupyter team released the JupyterHub web application to provide a platform where multiple users can login and access a Jupyter Notebook environment. When the number of users and memory requirements are low, it is easy to setup JupyterHub on a single server. However, setup becomes more complicated when we need to serve Jupyter Notebooks at scale to tens or hundreds of users.
In this paper we will present three strategies for deploying JupyterHub at scale on XSEDE resources. All options share the deployment of JupyterHub on a Virtual Machine on XSEDE Jetstream.
In the first scenario, JupyterHub connects to a supercomputer and launches a single node job on behalf of each user and proxies back the Notebook from the computing node back to the user's browser.
In the second scenario, implemented in the context of a XSEDE consultation for the IRIS consortium for Seismology, we deploy Docker in Swarm mode to coordinate many XSEDE Jetstream virtual machines to provide Notebooks with persistent storage and quota.
In the last scenario we install the Kubernetes containers orchestration framework on Jetstream to provide a fault-tolerant JupyterHub deployment with a distributed filesystem and capability to scale to thousands of users.
In the conclusion section we provide a link to step-by-step tutorials complete with all the necessary commands and configuration files to replicate these deployments.

\end{abstract}

%
%



\maketitle

\section{Introduction}

\subsection{Jupyter Notebooks and JupyterHub}

Jupyter Notebooks \cite{ipython2007,jupyter2016} are now part of the standard toolbox of researchers in any field of science, e.g. \href{https://github.com/dalya/WeirdestGalaxies}{astrophysicists analyzing spectra of millions of galaxies}\cite{sdss} or \href{https://github.com/theandygross/TCGA/tree/master/Analysis_Notebooks}{bioinformaticians working on head and neck cancer genomics}\cite{gross}.
Scientists choose Jupyter Notebooks because they provide a single environment where not only their code, but also their equations, figures, explanatory text and even interactive visualizations can be stored.
Moreover, the Jupyter Notebook web application provides an environment where they can prototype their algorithms or data analysis pipeline and iterate over those quickly and effectively. Once their work is complete, they have a persistent archive that can be reproduced in the future and easily shared, reused and remixed.

Although Jupyter Notebooks are a single-user web application, the Jupyter team also released JupyterHub, a multi-user web application. It provides an authentication mechanism and once the user logs in, it spawns a dedicated Jupyter Notebook single-user application and proxies it back to the user's browser.
JupyterHub is highly customizable, its main components are the Authenticator, e.g. \texttt{OAuthenticator} \footnote{\url{https://github.com/jupyterhub/oauthenticator}} that provides authentication through Google, Github or Globus, and the Spawner, e.g. \texttt{DockerSpawner} \footnote{\url{https://github.com/jupyterhub/dockerspawner}} that launches Jupyter Notebooks inside isolated Docker containers.

\subsection{Deployment of JupyterHub on XSEDE}

Deploying JupyterHub at scale to provide adequate computational resources to a large number of users is challenging because it requires a distributed infrastructure.
In this paper we will describe three different strategies where we exploit XSEDE \cite{xsede} resources to achieve this objective.

A foreseeable application of each of these deployments is to provide a common computational environment to students of a class, attendees of a workshop or members of a large scientific collaboration. In such cases, it is convenient that the users only need a web browser to access a pre-configured environment containing all the necessary software and data.
In the event that the deployment is persistent in time, the users also do not need to worry about software updates as the environment is centrally managed.

In each of the following scenarios, the users' Jupyter Notebooks are executed inside a custom container: Singularity for the deployment on HPC and Docker for the deployment on Jetstream \cite{jetstream2015}. The simpler option is to choose one of the containers provided by the Jupyter Team in the \href{jupyter/docker-stacks}{https://github.com/jupyter/docker-stacks} repository, for example the DataScience Notebook includes the Jupyter Notebook with Python, Julia and R kernels and many of the packages dedicated to data analysis and machine learning. Otherwise the JupyterHub deployment maintainers can leverage all the containers already publicly available on \href{DockerHub}{https://hub.docker.com/}, install additional software needed by their audience, then follow a few steps to add support for JupyterHub.

A second application, particularly relevant in the context of XSEDE, is to use one of these same strategies to deploy JupyterHub in conjunction with a Science Gateway. We will introduce Science Gateways and discuss the interactions with JupyterHub in the next section.

\subsection{JupyterHub and Science Gateways}

\begin{figure*}[ht]
\includegraphics[width=\textwidth]{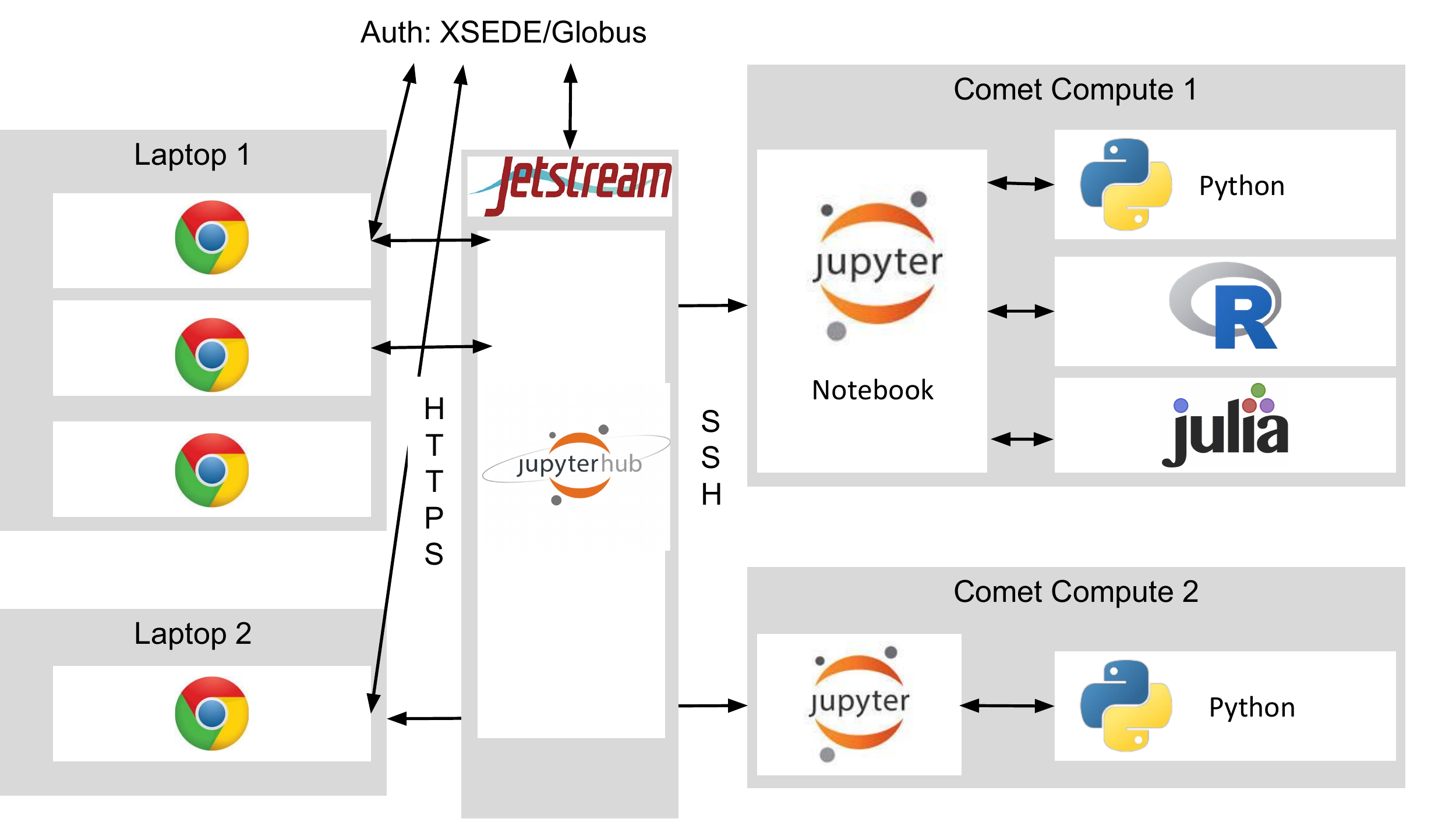}
\caption{JupyterHub on HPC - Multiple users, each with multiple browser tabs, can access, through HTTPS, the JupyterHub instance installed on a Jetstream Virtual Machine. They are redirected for Authentication via OAuth to XSEDE or Globus and then redirected back to JupyterHub. Then JupyterHub connects via SSH to Comet and launches a SLURM job. Once the job starts on a Comet computing node, it sets up a SSH tunnel back to JupyterHub to forward the Jupyter Notebook port and starts up a Jupyter Notebook inside a Singularity container. Each browser tab of the user interacts with a different kernel, possibly in any of the more than 50 different programming languages supported by Jupyter. \label{fig:jupyterhub_hpc}}
\end{figure*} 

\begin{figure*}[ht]
\includegraphics[width=\textwidth]{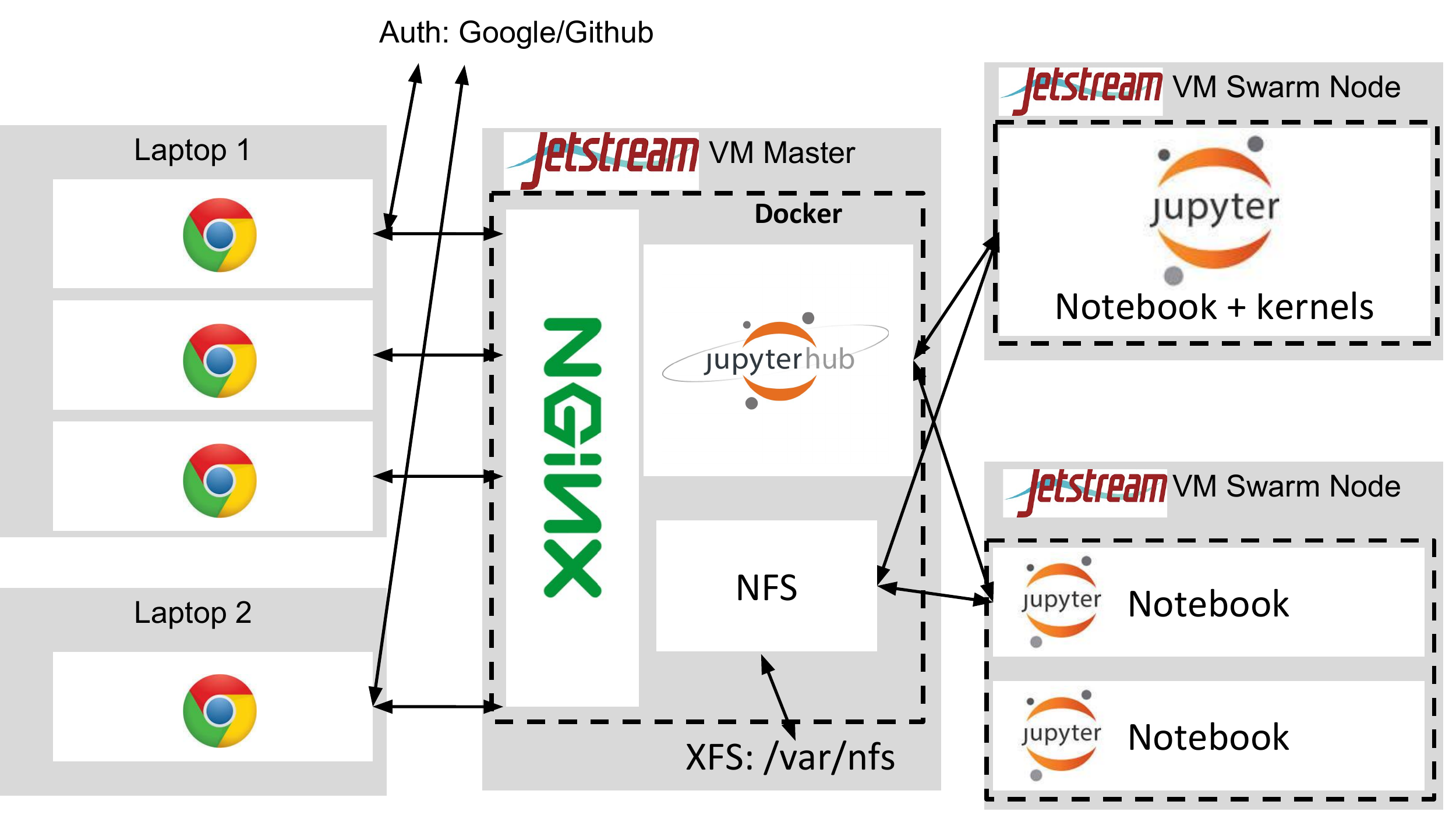}
\caption{JupyterHub on Jetstream with Docker Swarm - In this case, all applications run in isolated containers inside Docker (dashed lines). The NGINX web server and JupyterHub containers run on the master node, the other nodes execute Jupyter Notebooks inside containers. Docker Swarm handles internal networking automatically. The master node has a Jetstream volume attached which is formatted with XFS to support quotas. It is shared via NFS to provide persistent storage to all the nodes running users' containers.\label{fig:jupyterhub_docker}}
\end{figure*}  

\begin{figure*}[ht]
\includegraphics[width=\textwidth]{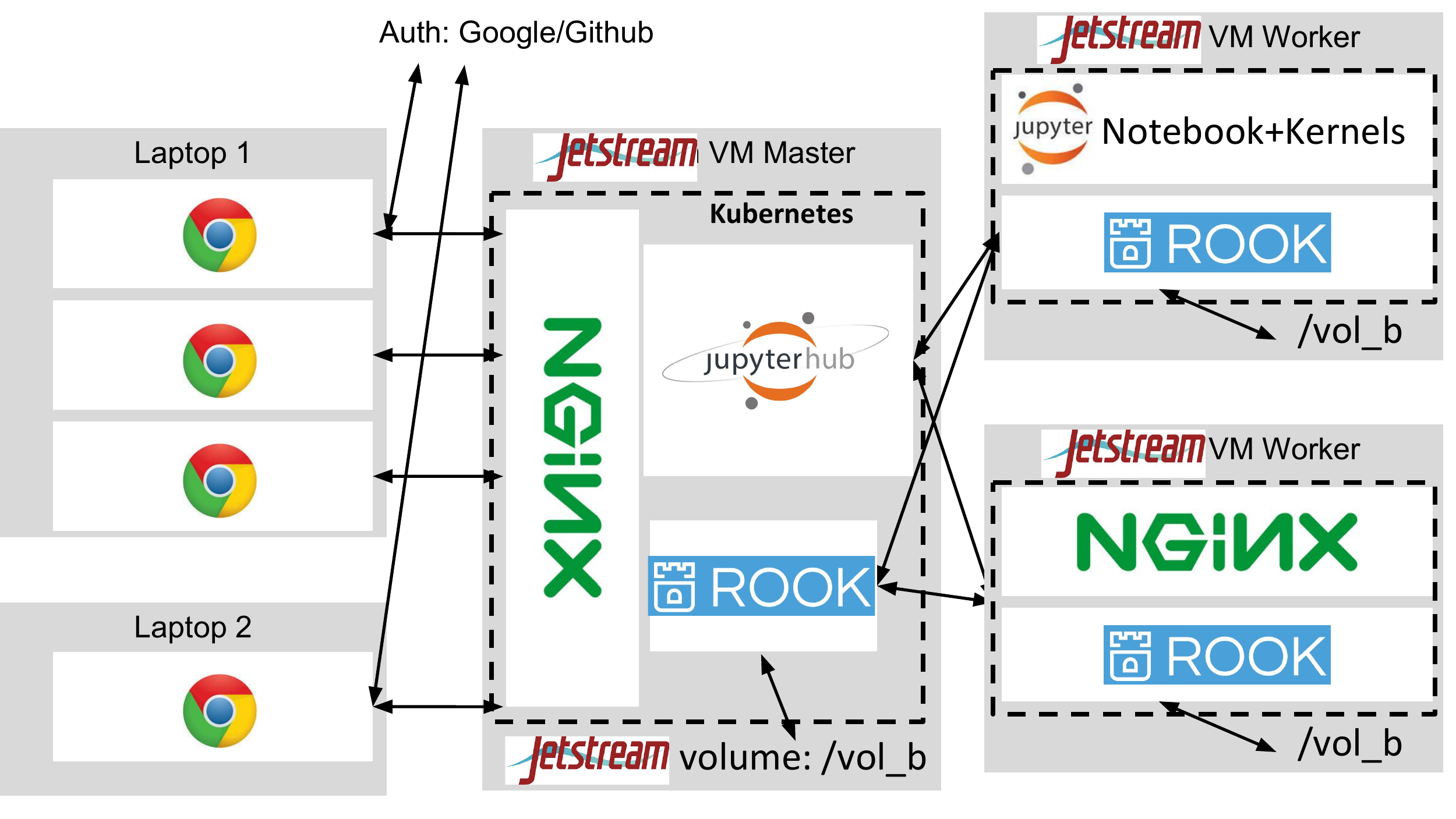}
\caption{JupyterHub on Jetstream with Kubernetes - JupyterHub runs as a container inside Kubernetes (dashed lines), the NGINX web server handles HTTPS and has two instances spread over the cluster of nodes, with routing handled automatically. JupyterHub sends a request to Kubernetes for a user container to be spawned each time a new user logs in. Persistent data storage is provided by Rook deploying the Ceph distributed file system across the Jetstream volumes mounted on all nodes. \label{fig:jupyterhub_kubernetes}}
\end{figure*} 

The world of High Performance Computing (HPC) is leveraging web technologies to lower the entry barrier to its computational resources. The culmination of this process is the birth of Science Gateways, websites where users can upload their data and/or configure a simulation pipeline. The computational work is sent in the background for execution on a supercomputer and the results made available through the same website for download or further processing.
A demonstration of the reach of Science Gateways is CIPRES \cite{cipres}, which provides phylogenetic tree inference to the science community and enabled over 3,700 publications since 2008, 753 of them in 2017 alone.
The National Science Foundation recognized the importance of these platforms by \href{https://www.nsf.gov/awardsearch/showAward?AWD_ID=1547611}{funding the Science Gateways Community Institute}\cite{sgci} to foster, incubate and support existing and new gateways.

JupyterHub shares some features with Science Gateways since they are both multi-user web platforms that provide computational services to scientists. In this section we provide example use cases of interaction between them.

The first scenario is where JupyterHub is deployed as a companion application to a Science Gateway. The scientist can launch computational jobs on XSEDE High Performance Computing clusters either through a web-based interface from the Science Gateway website or directly via its API from the Jupyter Notebook. Once the results are available, the scientist doesn't even need to download them, they can access a Jupyter Notebook running close to where data are stored. The Jupyter Notebook computing environment would have pre-installed all libraries needed to load, process and visualize the data. By using JupyterHub, Science Gateways not only provide computational power to execute the HPC job but also a convenient online environment for preparation of the inputs and post-processing of the outputs. In principle a scientist could proceed from the input data to publication-ready results without ever leaving the environment.

A more radical scenario is when JupyterHub itself is the Science Gateway. This kind of platform provides a Jupyter Notebook with a software environment customized for the target science community, continuously maintained and updated. Moreover, it can provide high-bandwidth access to data from XSEDE storage resources without the need to download data locally, even directly mounting those resources as local disks accessible transparently from the Notebook.

Regardless of the target application, in the next sections we will leverage XSEDE resources to deploy JupyterHub with a strategy that allows to serve Jupyter Notebooks with enough CPU, memory and disk to hundreds of users simultaneously.

\section{JupyterHub on HPC via the batch scheduler}

The first deployment strategy provides Jupyter Notebooks running on a compute node of a traditional HPC cluster by submitting jobs to the scheduler, for example SLURM \cite{slurm}.

One application of this kind of deployment is to ease access to supercomputing resources. Instead of accessing the resource through a terminal, users can just login to JupyterHub with their credentials and directly reach a compute node of a cluster. This has been deployed, for example, at the Minnesota Supercomputing Institute \cite{milligan}.

\subsection{Architecture of the deployment on Jetstream and Comet}
 
Figure \ref{fig:jupyterhub_hpc} provides an overview of the architecture of this deployment. First JupyterHub is deployed on a Ubuntu Virtual Machine on XSEDE Jetstream, for example using the Ansible recipe \footnote{\url{https://github.com/jupyterhub/jupyterhub-deploy-teaching}} by the Jupyter team.
Once the server is setup, it already has a JupyterHub instance with the web server NGINX \cite{nginx} sitting on top of it and handles HTTPS encryption with a Let's Encrypt certificate.
It is then possible to setup authentication with XSEDE credentials via the \texttt{CILogon} support implemented in the \texttt{OAuthenticator} project. Other options are Github, Google or Globus accounts.

Connection to the Comet supercomputer in our deployment is handled by the Spawner batchspawner\footnote{\url{https://github.com/jupyterhub/batchspawner}}. It has templates to interface with several schedulers like SLURM or PBS and manages connecting to the supercomputer and submitting a job.

An important detail is correctly choosing the account that is charged for the users' jobs. Science Gateways generally apply for a large allocation from XSEDE and receive a Community Account which is charged for all the jobs that scientists submit to the Gateway. One option is to follow this model and use the Community Account SSH key to submit jobs to the scheduler. Another option, available only if we authenticate the users through XSEDE credentials, is to request from CILogon an X509 token and authenticate to the supercomputer on behalf of the user with their credentials, so that each users' account is charged for the computational hours spent for the job.

The job duration, queue and any other configuration option can either be given upfront by the JupyterHub administrator for all users or we can allow the user to choose their preferences through a web-based form. The users could optionally choose a Docker or Singularity container to execute their Notebooks, so that they can completely customize their software environment.

After the job is in the queue, it is essential that the waiting time is short, since it wouldn't be useful if a user has to wait hours to start doing interactive work. Therefore, it might be necessary to provide a dedicated queue on the supercomputer with higher priority than regular batch jobs.

Once the job is assigned a compute node by the scheduler, it first sets up an encrypted tunnel back to the Jetstream Virtual Machine via SSH and then starts up a single-user Jupyter Notebook instance inside a custom Singularity container. JupyterHub has been waiting for the Jupyter Notebook to be reachable at a pre-defined port and can proxy it back to the users' browser.
The scientist never connects directly to the supercomputer compute node, which doesn't even need to have any open ports since all traffic goes through JupyterHub.

\section{JupyterHub on XSEDE Jetstream with Docker Swarm mode}

This section and the next focus on a deployment of JupyterHub on the XSEDE Jetstream cloud resource. Jetstream is an OpenStack deployment at Indiana University and the University of Texas at Austin that provides the users with on-demand Virtual Machines in a manner similar to Amazon Elastic Compute Cloud and Google Compute Engine.

This strategy, which is ideal for workshops, involves Docker Swarm and permanent storage provided by the traditional network file system NFS and is suitable for less than one hundred concurrent JupyterHub users on up to ten nodes. This setup was deployed during a Extended Collaborative Support Service (ECSS) \cite{ecss} consultation for Incorporated Research Institutions for Seismology (IRIS) to provide an easily accessible software platform for workshops. ECSS is a service provided to the principal investigators on XSEDE awards who would benefit from an assistance by XSEDE HPC or technology experts. ECSS collaborations are initially awarded for up to one year and may be renewed.

\subsection{Architecture of the deployment on Jetstream with Docker Swarm mode}

Figure \ref{fig:jupyterhub_docker} provides an overview of the architecture of this deployment.

The primary advantage of deploying on Jetstream is that we can quickly adjust the amount of resources available by adding and removing Jetstream instances to the setup.

Current versions of Docker include Swarm mode, which automatically manages a cluster of machines with Docker installed, manages deploying services across the nodes and ensures that the deployed services are always available. For example, if we have a web server container running in a node of the cluster and that node disappears for any reason, Docker automatically redeploys that container to another node and routes the traffic accordingly.

We start by setting up the master node with a recent version of Docker and activate Swarm mode. We can then have other nodes join the Swarm by running a specific Docker command and pointing to the master node. Therefore, we can test the setup first with just two nodes and then ramp up the number of nodes once we need to accommodate users.

The Jupyter team already provides a JupyterHub Docker container, therefore we can pull it to the machine and deploy it in the Swarm. Later we can customize its configuration by building the container ourselves to, for example, choose authentication via Google or Github accounts.

We also deploy the NGINX web server as a Docker Swarm service to handle HTTPS encryption, optionally using two instances of the NGINX container so we can share the load over two nodes. All necessary load-balancing and routing is transparently handled by Docker Swarm.

Once the user connects to the master node with their browser, their connection will be dispatched internally to one of the two NGINX containers that will handle HTTPS and proxy the JupyterHub container. The user will be redirected by the Authenticator to either Google or Github to login and then the JupyterHub container will run SwarmSpawner to launch a Jupyter Notebook Docker container somewhere on the Swarm. Once the container is initialized, the familiar Jupyter Notebook interface is proxied back to the user's browser. We can choose the Docker container either in the JupyterHub configuration or have the users choose their own, as long as it has the Jupyter Notebook installed and a script to interface it with JupyterHub.

Docker natively supports setting custom memory and CPU limits for each running container, so that the administrators can make sure a user is not slowing down all the other users on the same machine.

\subsection{Permanent storage with the network file system NFS}

On a distributed resource with no shared file system, permanent storage is a difficult problem to tackle. A user could have a working session on one node and then have the next working session on a different node, but would like their files available at all times.
Surprisingly, Swarm has no native support for this feature, but we can easily implement it using the traditional UNIX Network File System (NFS).

We can create a dedicated Volume through the Jetstream web interface Atmosphere and attach it to a node and format it with the XFS file system. This has the added bonus of supporting quotas on directories and it is our only option because all users run with the same UNIX account.

We can deploy an NFS container pinned to the node with the Openstack Volume and ensure that all other nodes in the Swarm can access the shared Volume. Finally, we can configure SwarmSpawner so that the first time a user requests a Jupyter Notebook, a Docker Volume is created as a folder in the NFS shared file system and mounted into the container. In subsequent sessions, wherever their Docker container happens to spawn in the Swarm, the same Docker Volume will be mounted on their container, thereby guaranteeing that the users always have access to their data.

Having defined quotas on the folders at the file system level, we know in advance what is the maximum number of users our infrastructure can support. We can also monitor the disk and email users when their quota is almost exhausted.

\section{JupyterHub on XSEDE Jetstream with Kubernetes}

The Jupyter team released a guide to deploy a scalable instance of JupyterHub, named Zero to JupyterHub\footnote{\url{https://zero-to-jupyterhub.readthedocs.io}}, on public cloud services, such as Google, Amazon and Microsoft, based on Kubernetes. Kubernetes is an open-source platform for orchestrating containers distributed on a cluster of nodes.

This deployment has been used in production at UC Berkeley with more than a thousand students for their introduction to Data Science class. The same recipe can be adapted to a private Kubernetes instance created on Jetstream.

Kubernetes is a powerful and complex piece of infrastructure, fortunately there are many resources available online to help with its steep learning curve.

\subsection{Architecture of the deployment on Jetstream with Kubernetes}

Figure \ref{fig:jupyterhub_kubernetes} provides an overview of the architecture of the deployment.

The first step is to install Kubernetes on Jetstream, the \texttt{data8} Data Science team released on Github a recipe to setup Kubernetes \footnote{\url{https://github.com/data-8/kubeadm-bootstrap}}, the \texttt{kubeadm} administration tool and the \texttt{Helm} Kubernetes package manager \footnote{\url{https://helm.sh}}. This is targeted at Ubuntu instances, therefore we create two Ubuntu Virtual Machines on Jetstream and install Kubernetes.

\subsubsection{Permanent storage with the distributed file system Ceph}

The Zero to Jupyterhub deployment supports permanent storage for the user's data. While public cloud services come pre-installed with storage providers for Kubernetes, on Jetstream we need to manually install and configure it.

We want to leverage Kubernetes itself to provide a storage solution that is distributed, scalable and fault-tolerant. One of the best available options in the Kubernetes Ecosystem is Rook \footnote{\url{https://rook.io}}, which simplifies the deployment and administration of the Ceph distributed file system \cite{ceph}.

Ceph federates disks from multiple machines in a network and stores data with redundancy so that if a node dies no data loss occurs. 

First we create Volumes for each node via the Jetstream Atmosphere interface and attach them to each of the Kubernetes instances. We then deploy Rook through Helm across our cluster and instruct it to use the Volumes across all nodes to create a single storage pool and setup a provider for Kubernetes.

\subsubsection{Install JupyterHub with Helm}

The Zero to JupyterHub project provides a Helm package to setup all the Docker containers on Kubernetes, which include two or more NGINX web servers and a JupyterHub container. It has a large set of configuration options and extensive documentation about them.

As usual we want to customize the Authenticator. Moreover, we need to instruct JupyterHub to use Rook for persistent storage of the user data and configure Kubernetes to expose the web ports of the NGINX container on the master node.

Once the user connects to the master node, it is redirected to HTTPS and routed to one of the nodes running the NGINX container which proxies the connection to the JupyterHub container that will handle authentication.
After the user is authenticated, JupyterHub, relying on \texttt{KubeSpawner}, requests that Kubernetes creates a new container for the user. It also sets limits on the amount of RAM, CPU and disk that the user is allowed to access.
As soon as the container starts, the Jupyter Notebook port is routed back to the JupyterHub container and then to the user.

\subsubsection{Scaling Kubernetes clusters up or down}

It is easy to add capacity to JupyterHub by creating more Virtual Machines on Jetstream and launching a single script to have them join the Kubernetes cluster.
Once they join the cluster, Kubernetes through Rook will prepare the additional storage space to be joined to the Ceph pool and the new user's container could be scheduled to run on this node.

If extra capacity is no longer needed, we can execute a Kubernetes command on the master node to schedule the migration of all services off of a specific node, then exclude it from the cluster and finally terminate the instance.

Ideally we would want to automatically scale the cluster size based on the load or the number of currently connected users. This is not currently available in this setup but it is in the road-map for future improvements.

\section{Conclusion}

We presented three different strategies for deploying JupyterHub that allow users to access XSEDE resources through a convenient and easy-to-use Jupyter Notebook interface. Step-by-step tutorials describing these strategies, all necessary commands and configuration files, and a recording of an ECSS webinar on this topic are available at \url{https://zonca.github.io/2017/12/ecss-symposium.html}.

The proposed recipes can be customized and improved: for example, in a deployment on Jetstream, instead of providing local storage, we could require XSEDE authentication and then use it to mount a XSEDE storage resource like Pylon or Oasis inside the user container, so that the user can access their data on HPC and JupyterHub.

The deployments described here could be automated for example with Ansible, however, those black-box recipes are more difficult to customize and to debug. A step-by-step tutorial allows the administrators to familiarize with each piece of infrastructure separately, customize more easily each step and identify quickly what is the problem if a failure occurs. In fact, the tutorials have optional steps where the administrators can perform dedicated functionality tests. 

In the future we plan to continue this work focusing on the Kubernetes deployment and provide more computational resources to the users that will allow them to launch a pool of other containers and \href{https://dask.pydata.org}{Dask} workers for distributed computing in Python.

\begin{acks}

We thank the following organizations and individuals for sharing their time and expertise:

\begin{itemize}
\item The Jupyter team, in particular Yuvi Panda, for providing a great software platform and a easy-to-use resource for deploying it and for direct support in debugging our issues
\item XSEDE Extended Collaborative Support Services for funding the work on deploying JupyterHub on Jetstream and for providing computational time on Jetstream
\item Pacific Research Platform, in particular, John Graham, Thomas DeFanti and Dmitry Mishin (SDSC), for access to their Kubernetes platform for testing
\item XSEDE Jetstream's Jeremy Fischer for prompt answers to questions about Jetstream
\end{itemize}

 The work is supported by the \grantsponsor{NSF}{National
    Science Foundation} under Grant
  No.:~\grantnum[https://www.nsf.gov/awardsearch/showAward?AWD_ID=1548562]{NSF}{1548562}
  \textit{XSEDE 2.0: Integrating, Enabling and Enhancing National Cyberinfrastructure with Expanding Community Involvement}.

\end{acks}

\bibliographystyle{ACM-Reference-Format}
\bibliography{sample-bibliography} 

\end{document}